\documentclass[debug,overfull]{epl} 
\usepackage{graphicx}
\usepackage{dcolumn}
\usepackage{amsmath}
\usepackage[latin1]{inputenc}
\begin{document} 
\title{Intermittency,   aging and  extremal fluctuations}
\shorttitle{Intermittency and aging } 
\author{Paolo Sibani\inst{1,2}\thanks{$^1$ Permanent address: Fysisk Institut, SDU,  Odense, DK}  
\and Henrik Jeldtoft Jensen\inst{2}}
\institute{\inst{1}Theoretical Physics, 1 Keble Rd, Oxford OX1 3NP, UK \\
\inst{2}Dept. of Mathematics,
Imperial College London, South Kensington Campus,  London SW7 2AZ, UK}  

\date{\today} 
\maketitle  
\begin{abstract}
    Aging in spin glasses is analyzed via the
    Probability Density Function  (PDF) of the heat transfer
    over a short  time $\delta t$ between system   and  heat bath.
    The PDF contains  a Gaussian part,
    describing reversible  fluctuations, and an exponential 'intermittent' tail. 
    We find that the relative weight of these two parts depends,   
     for fixed $\delta t$  on   the ratio of  
    the total sampling time $t$ to the age $t_w$.  
Fixing this ratio,  the intensity of the intermittent events
is proportional to $\delta t/ t_w$ and independent of the temperature.
    The Gaussian part has a variance with
    the same temperature dependence as the variance
    of the equilibrium energy in a system
    with  an exponential density of states.
    All  these   observations are  explained 
    assuming that for any $t_w$,  intermittent events
    are  triggered by local energy fluctuations 
    exceeding  those previously occurred.  
 \end{abstract}

 \section{Introduction} Aging   glassy systems undergo   reversible equilibrium-like 	
 	fluctuations alongside with  	irreversible configuration  changes. 
As  shown  experimentally in~\cite{Cipelletti03a,Buisson03,Buisson03a} 
for a number of different  cases,  these 	two aspects   can   
be disentangled by a statistical analysis of a  mesoscopic noisy 	 signal: 
	 within the probability density 	  function (PDF) of the   signal,  the  fluctuations 
         give  rise to a Gaussian peak, and the irreversible
        changes to  an asymmetric `intermittent' tail. This   probe of   aging dynamics  
         considerably  differs from    approaches   
	  utilizing    the  fluctuation-dissipation theorem, 
	and raises new theoretical issues    regarding e.g.\  the    statistical characterization 
 	  of  the    intermittent events
 	 and their  relationship  to  the   thermal fluctuations.  
\begin{figure}[ht!]
\twofigures[scale=0.37]{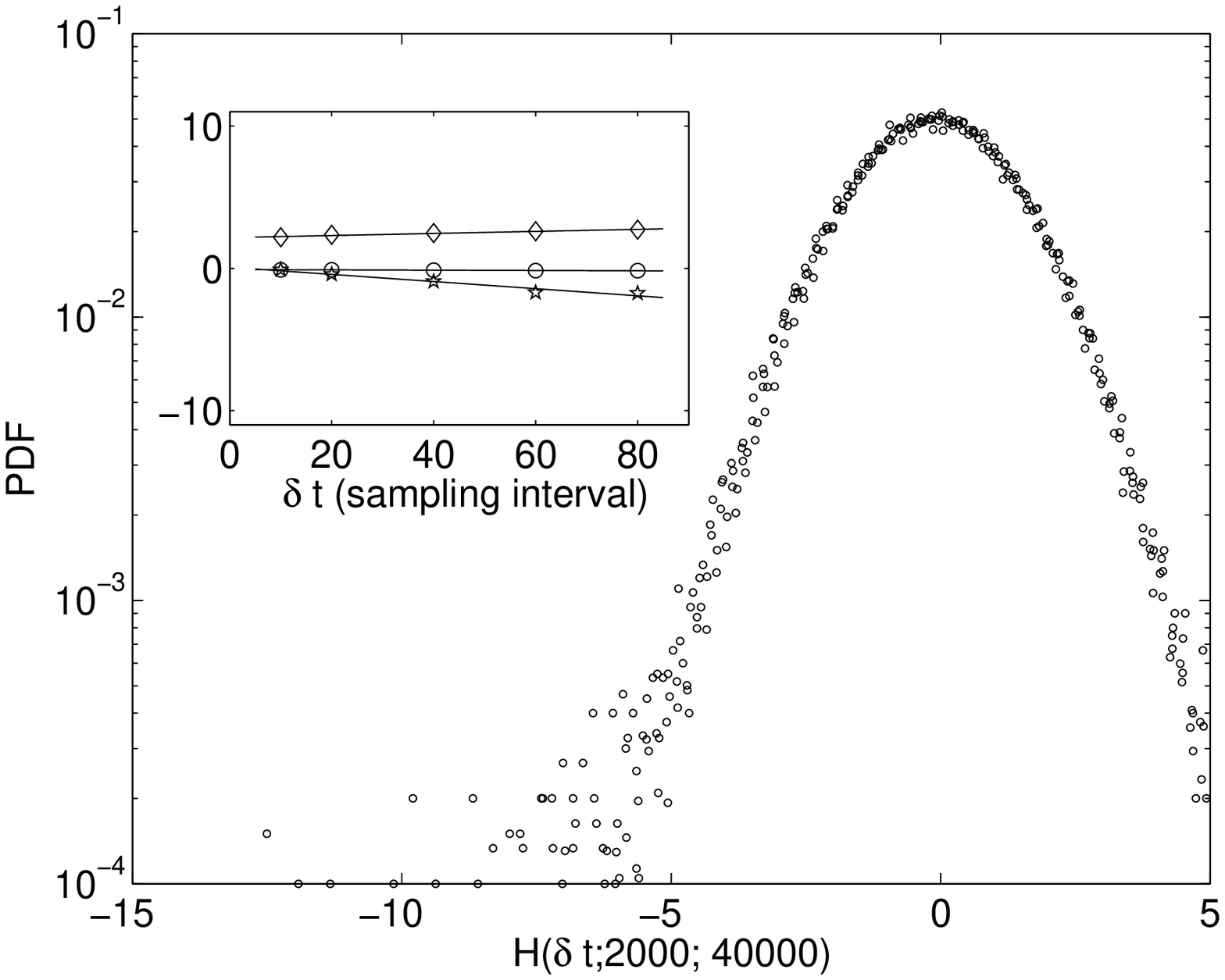}{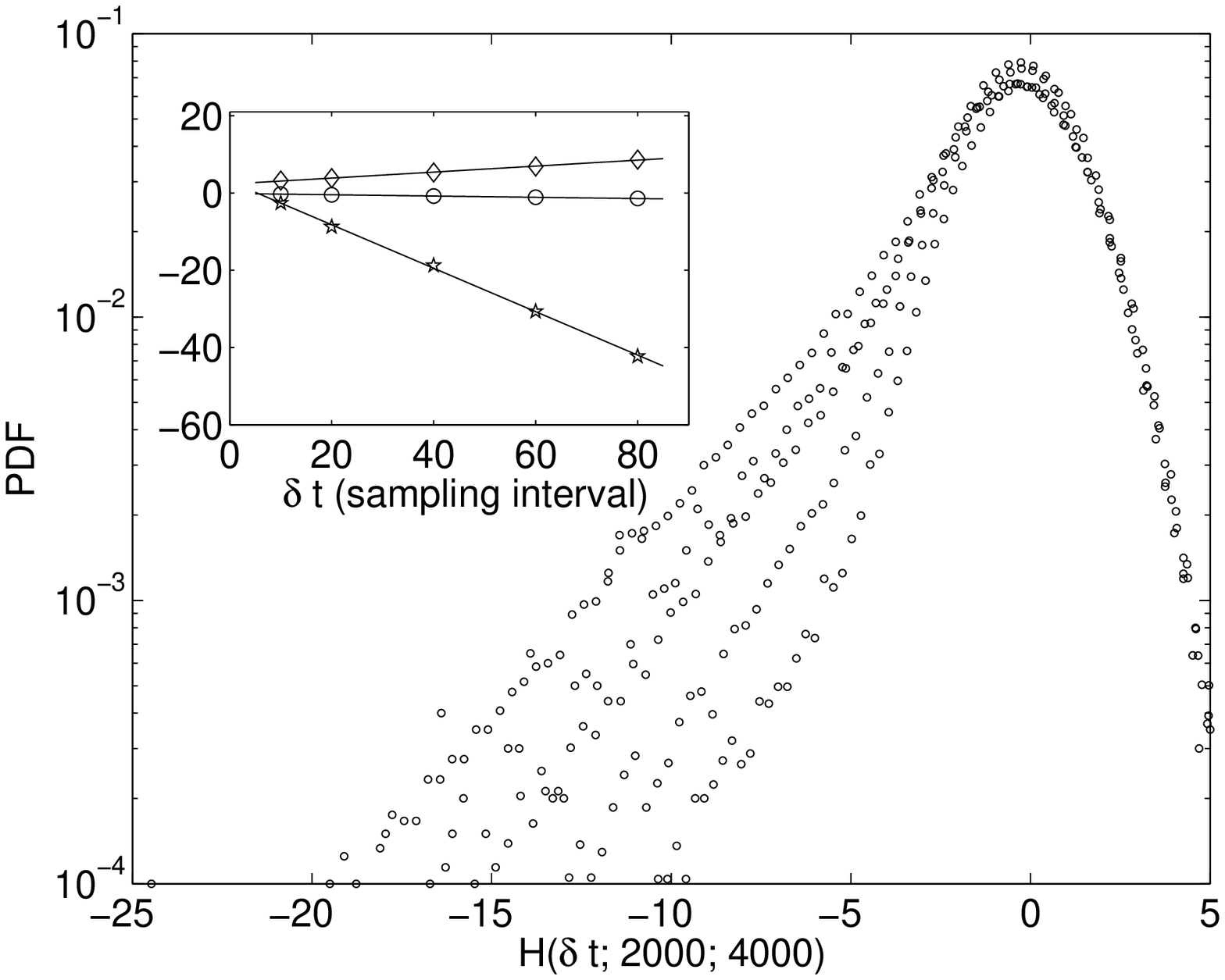} 
\caption{The PDF's  
(dots) of the heat transfer $H(\delta t, t, t_w)$,  sampled
in the observation window $[t_w, t_w+t]=[40000, 42000]$ 
for $\delta t= 10, 20, 40, 60$ and $80$. 
The  temperature is $T=0.3$.  
Insert:
the  mean (circles), 
variance (diamonds) and 
skewness (stars), are plotted versus the time difference
$\delta t$.}  
\label{equi} 
\caption{Data collected as
in Fig.~\ref{equi},   but with the 
observation window   $[t_w,t_w+t]=[2000,4000]$.
A strong asymmetric tail appears, producing  
 a  large negative   skewness when the observation 
time $t$ and the age $t_w$ are of comparable magnitude.     }
\label{nonequi} 
\end{figure} 
 We consider   the statistics of the heat-flow
 over a small interval $\delta t$.    
As was the case in~\cite{Cipelletti03a,Buisson03,Buisson03a}, 
 two  sub-processes  are discernible in the  PDF of the data:  
	  {\bf (i)} intermittent 
	events, referred to as \emph{quakes},
	 which appear in the left tail of the PDF.  These 
	irreversibly  release  the excess
	energy  trapped   by the initial  quench. 
        {\bf (ii)}  Gaussian fluctuations which occur 
        between the jumps of type {\bf (i)} and 
        correspond to reversible
	energy  fluctuations within  thermalized
	attractors.   Type {\bf (i)} events
         are  rare compared to   type {\bf (ii)}, but 
         are nevertheless   orders of magnitude    more frequent	
	than     Gaussian fluctuations of similar size. 
	Importantly,   the  rate of intermittent events
	decreases with the reciprocal of the age $t_w$  and is independent 
        of the  temperature $T$. Conversely, the Gaussian fluctuations
	are $t_w$ independent, but strongly $T$ dependent. 

	In this Letter  we identify the intermittent events with 
	the large and irreversible configuration changes of glassy
        systems  already described  in ref.~\cite{Sibani03},
	where they   were similarly called quakes. We hence borrow the idea  
	that their occurrence is due  to 
	energy fluctuations  larger than
 	any hitherto occurred, i.e.\  fluctuation records and show    
	that record-induced dynamics~\cite{Sibani93a,Sibani03} 
	predicts  the  key  dynamical features of the heat transfer PDF. 
	The  predictions    are first  spelled out and  tested
	 against simulation  data 
        for   a prototype glassy system, a three dimensional
	Ising  spin glass with short-ranged couplings.  
	 We then  argue in greater detail  for  the    
	 theoretical link between record-sized energy fluctuations, density of states
	within  metastable attractors  and intermittency, specifically 
	addressing  EA spin glasses and other models with short ranged
	interactions. We conclude with a  summary   and a brief  discussion.

  \section{Measurable  consequences of record-induced dynamics}  
 Non-equilibrium aging  is described in Ref.~\cite{Sibani03}
 in an  cartoon-like  fashion 
as a series of irreversible   `quakes' 
between dynamically inequivalent  metastable attractors
of  an energy landscape.  
   The  underlining physical   
assumptions  are: Firstly, that the stability of the
  'current attractor', as gauged by an 'exit' barrier, 
      increases by a tiny amount with each quake.
  Hence a \emph{positive}   energy fluctuation larger than all previous ones,
  i.e.\  a  record,    triggers  the    irreversible
  change of attractor. 
  Secondly, that  de-facto irreversibility of the quakes 
  arises due to the large   energy loss.
  I.e., crossing
  extremal barriers triggers irreversible configuration changes,
  while  smaller barriers are crossed reversibly.  
  Thirdly,    saddles are visited   rarely and in  a statistically
  independent fashion. 
 With these assumptions, the  quakes occurring in the interval $(t_w,t_w +t]$  
are well described  by  a  Poisson process with  logarithmic 
time arguments~\cite{Sibani03,Dall03}, and, by our identification,  so 
are the intermittent events. 
 
 The  average number of intermittent events is thus 
\begin{equation}
 \langle n_I(t_w,t_w+t)\rangle= \alpha \log(1+t/t_w), \quad \alpha > 0, 
\label{av_n}
\end{equation}
where  the parameter  $\alpha$  slowly   
  increases with the system size but 
  is  independent of $T$~\cite{Dall03}. 
Two   predictions which are   measurable 
from the PDF of intermittent data without access to  microscopic
information, are
easily  derived from Eq.~\ref{av_n}. 

Firstly, for  $\delta t \ll t_w$,  a single event occurs  in the interval  $\delta t$ with probability 
\begin{equation}
 p_{(t_w, t_w+  \delta t]}(1) =  \alpha  \frac{\delta t}{t_w} + {\cal O}(\delta t/t_w)^2,
\label{p_1}
\end{equation}
while multiple  events have zero probability
to  linear order in $\delta t/t_w$. Thus, 
the  rate of intermittent
events  decays as   $\alpha /t_w$ and is  independent of the
temperature.

Secondly,   the number of intermittent events
falling in an arbitrary time interval $t$ grows 
 on average  with $\log(1 + t/t_w)$.  As 
   reversible energy fluctuation occur at a constant
   rate, their number $n_G$ in the same
    interval   grows  linearly in $t$. 
 The  ratio $ n_I/n_G$   is hence given by
 \begin{equation}
   \frac{n_I}{n_G} \frac{t}{t_w}  \propto   \log(1 + t/t_w).
 \label{ratio}
 \end{equation} 
 	
	The  statistical properties  of records in a sequence of 
	  random numbers  are largely insensitive to   
	 the distribution from which these numbers are
	independently drawn~\cite{Sibani93a,Sibani03}. This  
	leads to  the predicted $T$ independence of the intermittency rate,
	which is however a   puzzling  conclusion
	when thermally activated barrier climbing 
	 arguably  produces  the attractor changes. 
	In the penultimate section of the Letter   we analyze
	this issue in some detail,  define the 
	 attractors more concretely for a  system with 
	short-ranged interactions, and derive 
	 Eq.~\ref{gauss_var}. The main result 
	is that the  cancellation of strong $T$ dependences
	needed to ensure consistency between record-dynamics
	and thermal barrier climbing  
	 occurs  precisely  if  the  energy fluctuations are
	drawn from  an exponential distribution. 
	In our case, this  distribution coincides  
	 with the   'local density of states' (LDOS) characterizing 
	   the energies available in local thermal equilibrium. 
	Writing the LDOS as 
  \begin{equation}
      {\cal D}(\epsilon)\propto \exp(\epsilon/\epsilon_0)  
      \label{expD}
      \end{equation} 
   produces  the heat capacity  
    \begin{equation}
    c_v(T)  \propto  \epsilon_0^2/(\epsilon_0 -T)^2, 
    \label{cv}
    \end{equation}
and leads to a    Gaussian variance  
of the heat transfer data
\begin{equation}    
   \sigma_G^2 \propto c_v(T)  T^2 \propto  \left( \frac{\epsilon_0 T}{\epsilon_0 -T} \right)^2, \quad T<\epsilon_0. 
   \label{gauss_var}
\end{equation} 
 This above  equation is our  third measurable  prediction.

 \begin{figure}[ht!]
\twofigures[scale=0.37]{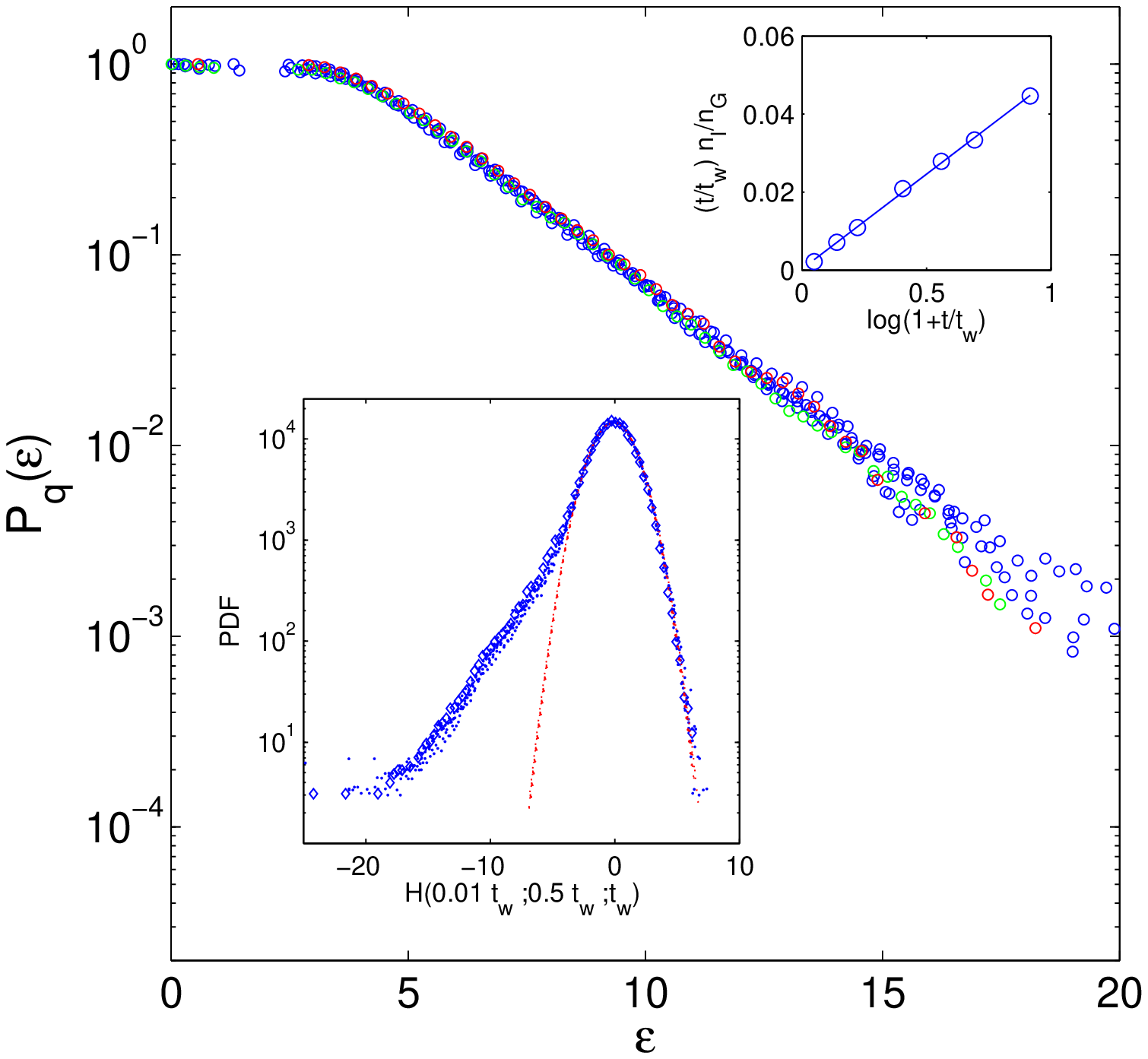}{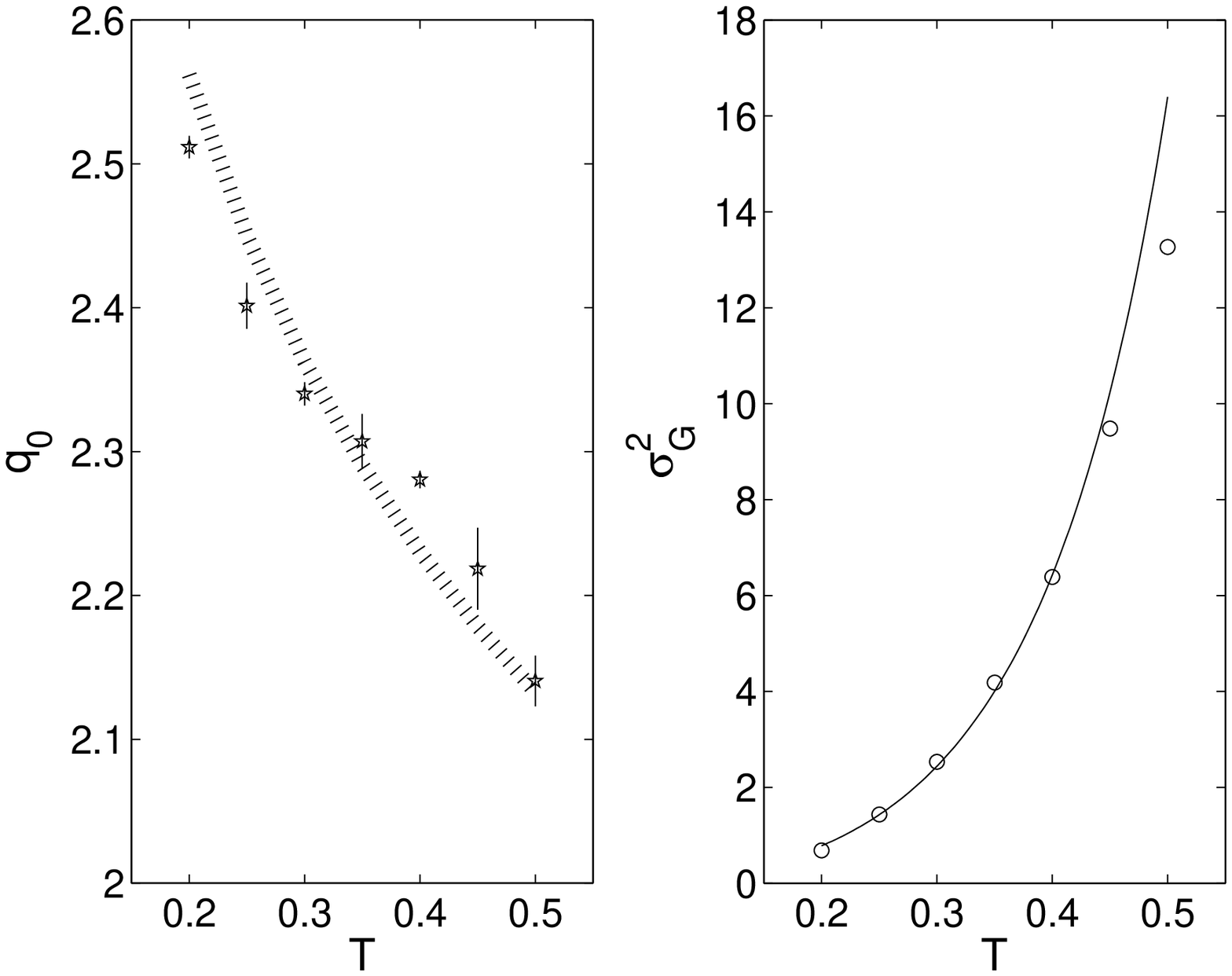} 
\caption{The probability $P_q(\epsilon)$ that  
the amount $q$ of energy intermittently released is larger than   
 $\epsilon$, is plotted versus $\epsilon$  
for  ages $t_w = 1000, 2000, 4000, \dots 64000$.
Lower insert: (unnormalized) PDF of the heat transfer $H(0.01 t_w, t_w/2,t_w)$.
The   full lines---indistinguishable 
on the scale of the plot---are fits   to 
Gaussians with zero average.
Upper insert: the relative weight $n_I/n_G$ (see text) of the intermittent
tail is obtained varying $t$ for fixed $\delta t$ 
 from seven values of $H(0.005 t_w, t,t_w)$
with $t_w=4000$ and $t=200,600, 1000,2000,3000,4000$ and $6000$.
The straight line is predicted by the theory.
The simulation temperature is  $T=0.3$.}  
\label{manypdf} 
\caption{Left  frame,  symbols:   reciprocal   slope
of  $ -\log(P_q(\epsilon))$ for large $\epsilon$, see Fig.~\ref{manypdf}, 
as a function of $T$. The dotted band 
is proportional to   $T^{-0.2}$.
Right frame, symbols:  variance $\sigma^2_G$ of the Gaussian
part of the PDF. Full line:
theoretical prediction $\sigma^2_G \propto T^2 c_v(T)$,  
with $c_v(T)$ given by Eq.~\ref{cv}.}
\label{moments} 
\end{figure}   
        
     \section{Simulation results}  The heat-flow data  stem from 
        simulations of   a standard $16^3$ 
	 Edwards-Anderson  spin-glass  with nearest neighbor 
	 Gaussian couplings~\cite{Edwards75}.  
     For speed we employ  an  
     event driven simulation technique~\cite{Dall01,Dall03}, 
	 whose  `intrinsic'   time   
	  corresponds,  for large systems, to 
        the  number of   Monte Carlo sweeps.
	The  Boltzmann constant is 
	set to one, and  age zero is   fixed  by
	the  instantaneous  deep thermal quench
	preceding the simulations.  
   
        We denote by  $E(t_w)$  the total  energy   at
        age   $t_w$, pick $\delta t < t_w$ and consider  a series of $n$ 
        contiguous time  intervals  of  length  $\delta t$.
        For the $k+1$'th interval, $k=0,1 \dots n-1$, the  heat $H$ exchanged  
        is calculated as  the  difference 
         $E(t_w+(k+1)\delta t) -E(t_w+k\delta t)$. Accordingly, $H$ is
	negative when    heat   leaves  the system.  
       Sampling $H$ over
	a total   observation time   $t=n \delta t$  produces 
	 the raw data behind the  Probability Density Functions (PDFs)
	discussed below. Since, as it turns out,  the  form of the PDFs  depends on  the 
	  three parameters 
         $\delta t,t$ and $t_w$,  the   heat transfer  
 	 is  denoted $H(\delta t,t,t_w)$. Omitting the 
	parametric time dependences,  the  PDF of $H$ is denoted by $p_H(x)$,
	whereas capital letters e.g.\ $P_q$ are used for cumulative distributions. 
	To improve the statistics simulations  are repeated  for  
	 625 or,  mainly,   1250   independent  runs for each 
	 set of physical parameters. 
         
	 Figures~\ref{equi} and \ref{nonequi} show data taken for 
	  $t=2000$,  $T=0.3$ and five  different  $\delta t$'s. 
	 In   Fig.\ref{equi}, $t_w=20 t$ while  $t_w=t$   
	in   Fig.~\ref{nonequi}. 
         In agreement with  previous  
 	observations~\cite{Cipelletti03a,Buisson03,Buisson03a,Crisanti04},
	the  PDF's have  a Gaussian part and an exponential tail.
         In  our figures,  this   tail  becomes   more  pronounced  as  $t \uparrow t_w$,   a  characteristic shift from 
	 equilibrium to  non-equilibrium dynamics
         usually studied through    correlations and 
	response functions~\cite{Alba87,Andersson92,Herisson02}. 
   
  The  asymmetry of the tail shows  that large negative 
  $H$ values are   associated   with irreversible energy losses.
  These events alone  carry the 
  linear $\delta t$  dependence of the mean, variance  and skewness
  shown in the inserts of Figs.~\ref{equi}~and~\ref{nonequi}, i.e.\ 
	the Gaussian part of the PDF has  always  zero average
	and a variance which depends  on $T$ but not on $\delta t$
or $t_w$, see e.g.\ Fig.~\ref{nonequi} (NB:  the scale differs from Fig.\ref{equi} ). 
	The   linear growth of the variance, also shown  in the
inserts,   continues   
   up to   $\delta t \approx  t_w$ (not shown) and 
  indicates   statistical  independence  of the  heat exchange  over consecutive 
  time intervals.

For fixed $t/t_w$ and increasing $t_w$, the rate
of intermittent activity decreases  as $1/t_w$ in a 
temperature \emph{independent} 	  fashion. 
This   $1/t_w$ scaling is shown for  $T=0.3$   
in  Fig.~\ref{manypdf}, where a  64-fold increase  of $t_w$ 
is offset by  choosing  $\delta t= t_w \delta_0$ 
with  $\delta_0=1/100$. 
The seven    $t_w$ values  considered are 
 $1000 \cdot 2^n$, for $n=0,1,\ldots 6$. 
 
The good agreement with the prediction given in
Eq.~\ref{ratio} is    demonstrated  in the upper insert of Fig.~\ref{manypdf}
by plotting $(n_I/n_G)(t/t_w)$ versus $\log(1 + t/t_w)$.  
For fixed $\delta t = 20$ and  $t_w=4000$, we consider 
observation times  $t=200,600,1000,2000,3000,4000$ and $10000$ (circles).
The  line is  a linear fit to the data.
The   unnormalized 
PDF's of $H(t_w/100, t_w/2,t_w)$ (dots) and  the respective    
   Gaussian fits (overlapping lines) are plotted in the lower insert.  
The   main panel  shows the `tail'  distributions   $P_q(\epsilon)$.   
 For large negative $H$ values these    
 describe the  difference $q$
 between   the empirical PDF value 
and the 
 Gaussian  probability for an energy  fluctuation
of size  $H(t_w \delta_0, t_w/2,t_w)$. The latter
is estimated   by  fitting a  Gaussian of zero 
average  to   the `central part' of the PDF, 
 as  delimited  by  a symmetric cutoff at     $\approx \pm 2 \sigma_G$. 
$P_q(\epsilon)$  is later interpreted as  the probability that a single 
 quake  releases an  energy $q$  larger than 
$\epsilon$.
   The size distribution of the quakes is nearly   
  exponential,   with a decay constant $1/q_0$.
  The   $\delta t/ t_w$ scaling   
  procedure   yielding     Fig.~\ref{manypdf}
was repeated for  $T=0.2,   0.4 $ and $0.5$,  
 with the same result.
 Additionally, for  $T=0.25,0.35, 0.45$,  six 
 sets of simulations  were  carried out 
 at fixed     $t_w = 16000$. 
 For each  $T$, averaging  over the outcomes  yields the  
 $q_0$  and  $\sigma^2_G$ values plotted  
 in  Fig.~\ref{moments} as circles. 
 The  full line in the right panel of Fig.~\ref{moments}
 is obtained by  Eqs.~\ref{gauss_var}~and~\ref{cv},
 using  a  fitted  pre-factor, which amounts to a vertical
parallel shift of the theoretical line.
  The value  $\epsilon_0=0.86$   used in Eq.~\ref{cv}
  is  very  close to  reported 
 values of the critical temperature in 
 \emph{3d} Gaussian spin glasses~\cite{Young98}.

 We note that while the    statistical variation of  $\sigma^2_G$ over different
 simulations set  is  insignificant,
 $q_0$ is estimated   with decreasing  accuracy
 as $T$ and $\sigma_G$  increase: 
  the distinction between 
     quakes and fluctuations is   moot
     for    $\epsilon$ smaller than  $\approx 2  \sigma_G$,
     and the  undetermined 
 part of the distribution  must be excluded from the fit. 
 With the present numerical effort and accuracy, we can only analyze 
 data below $T=0.6$.  
 The   choice of cut-off gives a systematic 
  error, not included in  the  $\pm \sigma$ error bars.  
     The gray  band  shows   the qualitative agreement with the  $T^{-0.2}$ scaling of the
 energy difference  between   local minima  found in Ref.~\cite{Dall03}.
 This temperature dependence is rather weak compared to the
$T$ dependence of  $\sigma^2_G$. 

 \section{Record fluctuations and intermittency}   
In extended systems with short ranged correlations,
thermal fluctuations    are localized  within well separated sub-volumes
whose linear size is of the order of the thermal 
correlation length\cite{Kisker96,Berthier02}.  
 As e.g.  suggested in ref~\cite{Bertin03}, each  
 thermalized sub-volume can   be treated as a   small glassy 
system,  with 
 a `local'  energy landscape.   Such landscapes have numerous 
  energy  minima. For     each  minimum $x$,  
  a metastable attractor can be defined as a  configuration
  space neighborhood, e.g.\  
  the set of configurations connected to  $x$
 by  paths requiring  energy changes
 below  a given  threshold~\cite{Sibani93}.  
 The      `local density of states' is the 
 density of state (DOS) restricted to  the 
 configurations belonging to the attractor.
LDOS    for several glassy
systems~\cite{Sibani93}   are indeed  
 well approximated by exponential functions of  the energy  
$\epsilon$ measured from the lowest minimum of the attractor.
The    divergence of the   heat capacity seen in Eq.~\ref{cv} 
at $T = \epsilon_0$  
implies that  for $T \geq \epsilon_0$
the thermal stability of the attractor vanishes. In practice
a  rounding  off occurs 
 once the weight   of  configurations
outside the attractor becomes important.  
 
       The Arrhenius time
     belonging to the typical extremal barrier crossed
     at time scale $t_w$ is obtained  as 
     follows: first we note that  
     the  Boltzmann  distribution
	corresponding to an  exponential LDOS is
     itself exponential, and given by 
     $P_E(b) = \exp(-b a(T))$, where         
     $a(T) = (1/T -1/\epsilon_0)$ is an 
      effective inverse temperature.
     Secondly,  with a constant microscopic attempt rate, 
      the largest   fluctuation 
     on a  time scale $t_w$ is  the largest   
     of  ${\cal O}(t_w)$ values drawn independently from  
        $P_E$.  
    Applying a  well known mathematical result~\cite{Leadbetter83},
       $b_{max}(T,t_w) \propto \ln(t_w)/ a(T)$. 
    Hence,   $a(T)$ cancels, leading to the $T$ independent 
      Arrhenius time   $\tau_A = \exp(a(T) b_{max}) = t_w$. 
     A small negative temperature  shift  imposed 
	at $t_w$  generates a mismatch between the  extremal barrier 
	  previously established  and  the thermal fluctuations
	  available to cross it. The  
	  Arrhenius time and the intermittency rate
	  acquire  then  a  peculiar temperature 
	\emph{dependence}  which has recently been
      verified in simulations~\cite{Sibani04a}.

 Clearly,   the sum of   many independent
 energy fluctuations  
is   a Gaussian quantity\footnote{Using a  very  small  $\delta t$ 
 the assumption of independence breaks down, and the  
Gaussian is  replaced by a back-to back double exponential.}, and so is  
  the central part of the heat transfer PDF,  
which arises  from  differences between  equilibrium  
energy  fluctuations. The Gaussian   variance   $\sigma^2_G$ 
 is   therefore  twice the 
equilibrium variance of the energy. The latter 
follows from Eq.~\ref{expD}
and  textbook arguments  of equilibrium statistical 
mechanics.

Finally, figs.\ref{equi}-\ref{manypdf}   show 
how the   record  dynamics scenario 
accounts for the dependence  of  the heat transfer   PDF, 
 on $\delta t, t$ and $t_w$. Writing the latter  as a a sum
of a Gaussian and an intermittent part, $p_H \propto p_G + p_I$, 
the tail has  the form
\begin{equation}
p_I(x) \propto \log(1+t/t_w) \frac{\delta t}{t_w} p_q(-x).
\end{equation} 
The  factor $\frac{\delta t }{t_w}$ combines  the linear   $\delta t$
dependence of the first three moments of $H$  seen    
in Figs.\ref{equi}~and~\ref{nonequi} with 
 the $\delta t/t_w$ scaling of the intermittency PDF data
shown in Fig.~\ref{manypdf}, for fixed $t$.
The interpretation of  $p_q(-x)$  as  the  
  PDF of the heat transferred in a single quake
hinges on  $\delta t/t_w$ being  proportional to  the   probability
of a single intermittent event, see Eq.\ref{p_1}.  

    An exponential  $p_q$ 
    could  arise quite  simply: the  sequence of  
   mainly downhill moves constituting the quake    
    releases  an   exponentially  distributed  energy 
    if each    move can be   the last
   with equal   probability.  
    Lacking  a   prediction  for the  weak  $T$ dependence 
    of the  scale parameter  $q_0$, we just note its   qualitative 
    agreement with  the  average energy difference
    between deep minima of the spin-glass landscape~\cite{Dall03}.

\section{Discussion}   
The correspondence between   intermittent events  
and  attractor  changes  was previously  
 investigated in ref.\cite{Crisanti04},   
 using thermal quenches into inherent structures.
We similarly link  intermittency 
 to   quakes, and derive  and verify  novel   analytical predictions     
for   the    heat transfer PDF, based on the assumption that 
  record-sized positive energy fluctuations within the attractors
  induce the quakes. 
 Record-induced dynamics  has  generic  implications  e.g.:  
 the $\delta t/t_w$   scaling form of the PDF  
 exemplifies  the  decelerating dynamics found  
besides spin-glasses~\cite{Andersson92,Herisson02,Berthier02,picco01}  
 in   glasses~\cite{kob00b}, 
driven dissipative systems\cite{Sibani01,nicodemi01}, 
and   evolution models~\cite{Sibani99a,Anderson04}.
Nearly exponential densities  of states  
are widespread~\cite{Sibani02,Jain02},
and  were used already  two decades ago~\cite{grossmann85}
in mesoscopic models of glasses.
We thus expect our  analysis  to    broadly apply  to glassy systems
whenever reversible fluctuations and   irreversible
quakes are clearly separated, i.e.   at sufficiently  low temperature.  
With this limitation,    it should also be    extensible 
to other types of measurements,  e.g.\  the fluctuations of 
two-point correlation functions~\cite{Cipelletti03a}. 
Generalizations   are   needed  to  account for   
small but systematic deviations from $t/t_w$   scaling, i.e. sub-aging, 
whose origin is currently debated~\cite{rodriguez02}. 
  
Extremal (Gumbel) distributions  are 
ubiquitous, e.g.\  Chamon et al.~\cite{Chamon04}
finds them to describe    local correlators in  aging systems,
including  systems without quenched disorder, and without local energy
minima. 
 Since  the heat absorbed by an extremal equilibrium fluctuation 
is  masked by the  much larger amount of heat   
released in the following quake,  the distribution of 
extremal energy fluctuations is   not directly 
accessible through  intermittent heat-transfer data. 
The relation with  effective
 temperature descriptions of fluctuation-dissipation data~\cite{Cugliandolo97,Calabrese04} 
  warrants a separate discussion. Here we only   note 
  that   these descriptions  concern 
 the   asymptotic limit of the fluctuation-dissipation
 ratio for large times. This  regime differs  from the  one presently considered  
 and is  not  always reached  experimentally~\cite{Buisson03,Herisson02} 

    {\it Acknowledgments-.} P.S. thanks J. Dall for his
    interest, and Pasquale
    Calabrese,   Sergio Ciliberto, Alan Parker, David Sherrington and 
    Veronique Trappe for useful discussions. Financial support from
    the Danish SNF and from EPSRC  is gratefully acknowledged. 
  \bibliographystyle{unsrt}

\bibliography{SD-meld} 

\begin{thebibliography}{10}

\bibitem{Cipelletti03a}
Luca Cipelletti, H.~Bissig, V.~Trappe, P.~Ballesta \and S.~Mazoyer.
\newblock {\em J. Phys.:Condens. Matter}, 15:S257--S262, 2003.

\bibitem{Buisson03}
L.~Buisson, L. Bellon and S.~Ciliberto.
\newblock {\em J. Phys. Cond. Mat.}, 15:S1163, 2003.

\bibitem{Buisson03a}
L.~Buisson, S.~Ciliberto and A. Garciamartin.
\newblock {\em Europhys. Lett.}, 63:603, 2003.

\bibitem{Sibani03}
Paolo Sibani and Jesper Dall.
\newblock {\em Europhys. Lett.}, 64:8--14, 2003.

\bibitem{Sibani93a}
P.~Sibani and Peter~B. Littlewood.
\newblock {\em Phys. Rev. Lett.}, 71:1482--1485, 1993.

\bibitem{Edwards75}
S.~F. Edwards and P.~W. Anderson.
\newblock {\em J. Phys. F}, 5:965--974, 1975.

\bibitem{Dall01}
Jesper Dall and Paolo Sibani.
\newblock {\em Comp. Phys. Comm.}, 141:260--267, 2001.

\bibitem{Dall03}
Jesper Dall and Paolo Sibani.
\newblock {\em Eur. Phys. J. B}, 36:233--243, 2003.

\bibitem{Crisanti04}
A.~Crisanti and F.~Ritort.
\newblock {\em Europhys. Lett.}, 66:253--259, 2004.

\bibitem{Alba87}
M.~Alba, J.~Hammann, M.~Ocio, and Ph. Refregier.
\newblock {\em J. of Appl. Phys.}, 61:3683--3688, 1987.

\bibitem{Andersson92}
J-O. Andersson, J.~Mattsson  and P.~Svedlindh.
\newblock {\em Phys. Rev. B}, 46:8297--8304, 1992.

\bibitem{Herisson02}
D.~H{\'{e}}risson and M.~Ocio.
\newblock {\em Phys. Rev. Lett.}, 88:257202, 2002.

\bibitem{Young98}
A.~P. Young, editor.
\newblock {\em Spin glasses and random fields}.
\newblock World Scientific, Singapore, New Jersey, London, Hong Kong, 1998.

\bibitem{Kisker96}
J.~Kisker, L.~Santen, M.~Schreckenberg, and H.~Rieger.
\newblock {\em Phys. Rev. B}, 53:6418--6428, 1996.

\bibitem{Berthier02}
Ludovic Berthier and Jean-Philippe Bouchaud.
\newblock {\em Phys. Rev. B}, 66:054404, 2002.

\bibitem{Bertin03}
J-M.~Drouffe E.M.~Bertin, J-P.~Bouchaud and C.~{Godr\`{e}che}.
\newblock {\em J. Phys. A}, 36:10701--10719, 2003.

\bibitem{Sibani93}
P.~Sibani, C.~Sch{\"{o}}n, P.~Salamon  and J.-O. Andersson.
\newblock {\em Europhys. Lett.}, 22:479--485, 1993.

\bibitem{Leadbetter83}
M.~R. Leadbetter, Georg Lindgren, and Holger Rotzen.
\newblock {\em Extremes and related properties of random sequences and
  processes}.
\newblock Springer-Verlag, New York Heidelberg Berlin, 1983.

\bibitem{Sibani04a}
Paolo Sibani and Henrik~Jeldtoft Jensen. 
\newblock {\em JSTAT}, page P10013, 2004.

\bibitem{picco01}
F.~Ricci-{T}ersenghi M.~Picco and F.~Ritort. 
\newblock {\em Eur.Phys.J. B}, 21:211--217, 2001.

\bibitem{kob00b}
W.~Kob and {J.L. Barrat}. 
\newblock {\em European Physical Journal B}, 13:319--333, 2000.

\bibitem{Sibani01}
P.~Sibani and C.~M. Andersen. 
\newblock {\em Phys. Rev. E}, 64:021103, 2001.

\bibitem{nicodemi01}
Mario Nicodemi and Henrik~Jeldtoft Jensen.
\newblock {\em J. Phys A}, 34:8425, 2001.

\bibitem{Sibani99a}
P.~Sibani and A.~Pedersen. 
\newblock {\em Europhys. Lett.}, 48:346--352, 1999.

\bibitem{Anderson04}
Paul~Anderson, Henrik Jeldtoft~Jensen, L.P.~Oliveira  and Paolo Sibani. 
\newblock {\em Complexity}, 10:49--56, 2004.

\bibitem{Sibani02}
P.~Sibani and J.~C. Sch{\"{o}}n. 
\newblock In J.~Fagerholm, editor, {\em Applied Parallel Computing, Lecture
  Series in Computer Science, Springer Verlag}, 2002.

\bibitem{Jain02}
Tushar~S. Jain and Juan~J. dePablo. 
\newblock {\em J. Chem. Phys.}, 116:7238, 2002.

\bibitem{grossmann85}
S.~Grossmann, F.~Wegner, and K.~H. Hoffmann. 
\newblock {\em J. Physique Letters}, 46:575--583, 1985.

\bibitem{rodriguez02}
G.~F. Rodriguez, G.~G. Kenning, and R.~Orbach. 
\newblock {\em Phys. Rev. Lett.}, 91:037203, 2002.

\bibitem{Chamon04}
{C.~Chamon, P.~Charbonneau, L.~F.~Cugliandolo, D.~R.~Reichman and M.~Sellitto}. 
\newblock {\em J. Chem. Phys.}, 121:10120, 2004.

\bibitem{Cugliandolo97}
Leticia F.~Cugliandolo \and Jorge Kurchan~\and Luca~Peliti. 
\newblock {\em Phys. Rev. E}, 55:3898--3914, 1997.

\bibitem{Calabrese04}
{P.~Calabrese and A.~Gambassi}. 
\newblock {\em cond-mat0410357}, 2004.

\end{thebibliography}
 
\end{document}